\documentclass[aps,prd,showpacs,eqsecnum,floatfix]{revtex4}

\usepackage{graphicx}
\usepackage{dcolumn}
\usepackage{bm}

\begin{document}


\title{Determination of the internal structure of neutron stars\\
from gravitational wave spectra}

\author{L.K. Tsui}
\author{P.T. Leung\footnote{Email:
ptleung@phy.cuhk.edu.hk}}
\author{J. Wu}

\affiliation{%
Physics Department and Institute of Theoretical Physics, The
Chinese University of Hong Kong, Shatin, Hong Kong SAR, China.
}%

\date{\today}
\def\tomega{ \tilde{\omega} }
\def\tOmega{ \tilde{\Omega} }
\def\tr{ \tilde{r} }
\def\tx{ \tilde{r}_* }
\def\tV{ \tilde{V} }
\def\tpsi{ \tilde{\psi} }
\def\trho{ \tilde{\rho} }
\def\tP{ \tilde{P} }
\def\tR{ \tilde{R} }
\def\tX{ \tilde{R}_* }
\def\tm{ \tilde{m} }
\def\tphi{ \tilde{\phi} }
\def\tnu{ \tilde{\nu} }
\def\tlam{ \tilde{\lambda}}
\def\tepsilon { \tilde{\epsilon}}
\def\tJ { \tilde{J} }
\def\tU{ \tilde{U} }
\def\cc{{\cal C}}
\def\a{ {\rm a}}
\def\c{ {\rm c}}
\def\e{ {\rm e}}
\def\p{ {\rm p}}
\def\f{ {\rm f}}
\def\d{ {\rm d}}
\def\i{ {\rm i}}
\def\r{ {\rm r}}
\def\rr{ {\rm r} }
\def\ri{ {\rm i} }
\begin{abstract}
In this paper the internal structure of a neutron star is shown to
be inferrable from its gravitational-wave spectrum. Iteratively
applying the inverse scheme of the scaled coordinate logarithmic
perturbation method for neutron stars proposed by Tsui and Leung
[Astrophys. J. {\bf 631}, 495 (2005)], we are able to determine
the mass, the radius and the mass distribution of a star from its
quasi-normal mode frequencies of stellar pulsation. In addition,
accurate equation of state of nuclear matter can be obtained from
such inversion scheme. Explicit formulas for the case of axial
$w$-mode oscillation are derived here and numerical results for
neutron stars characterized by different equations of state are
shown.
\end{abstract}

\pacs{04.40.Dg, 04.30.Db, 97.60.Jd, 95.30.Sf}
\keywords{}
\maketitle

\section{Introduction}
As remnants of supernova explosions, neutron stars  comprised of
matters with subnuclear and supranuclear densities have long been
a major subject of intense interest for researchers in
astrophysics, nuclear physics and particle physics since the
pioneering studies of Oppenheimer and Volkoff
\cite{Oppenheimer:1939ne} and Tolman \cite{Tolman:1939jz}. Their
internal structure is likely to provide a direct test for theories
of nuclear matter, quark matter and high energy physics (see e.g.
\cite{ComStar,APR,Lattimer:2001} and references therein). For
example, through measurement of the Eddington flux and the
redshift in the X-ray burst, the mass and the radius of the
neutron star EXO 0748-676 were simultaneous determined, and
relevant data were used to rule out soft equations of state (EOS)
for nuclear matter \cite{Ozel,Cottam}. Likewise, a few
astronomical observations revealing evidences for the existence of
the strange quark star (SQS), a variant of the neutron star (see
e.g. \citep{SQS_1,SQS_2,Cheng}), to some extent resuscitate the
theory of quark matter \citep{MITBM,Witten,AFO,PBP}.

On the other hand, the huge gravitational field accompanying a
dense neutron star also attracts the attention of relativists. In
particular, undulating neutron stars are expected to be promising
sources of gravitational waves. For instance, when
gravitational-wave detectors of various designs are in full swing
in one or two decades (see e.g. \citep{Hughes_03,Grishchuk} and
references therein), the frequency of detection of gravitational
waves emitted in the mergers of binary neutron stars could be as
high as several hundreds per year
\citep{Belczynski:2001uc,Hughes_03,Kalogera_04}. Besides,
gravitational waves emitted during asymmetric stellar core
collapses leading to the formation of neutron stars could also be
detectable \citep{Lindblom:1998wf,Fryer:2001zw}. It is then likely
that neutron stars could be surveyed in the gravitational wave
channel. More importantly, gravitational waves can easily
penetrate the core of a neutron star without being absorbed, and
might carry useful information about the deep interior of the
star. As the behavior of nuclear matter at high densities is not
yet clearly known, the internal structure of neutron stars is
masked and often gives rise to many debatable issues. Hence, a lot
of in-depth studies have been sparked off by the possibility of
unveiling the internal structure of neutron stars from their
gravitational-wave signals. Detailed examination of such signals,
which has been coined ``gravitational-wave asteroseismology", were
carried out by various groups of researchers (see e.g.
\citep{Andersson1996,Andersson1998,Ferrari,Kokkotas_2001,super_prl,
Kojima_02,Ferrari2003MNRAS,Sotani_03,Ferrari2004}) to study
effects of different physical processes, including superfluidity,
occurrence of quark matter, and phase transitions.

Gravitational waves emitted from a neutron star  are commonly
analyzed in terms of quasi-normal modes (QNMs) characterized by
complex eigenfrequencies $\omega=\omega_r-i\omega_i$, with
$\omega_r$ and $\omega_i$ measuring respectively the rates of
pulsation and damping \citep{rmp,kokkotas_rev}. To infer the
physical parameters of a neutron star from its QNM frequencies,
Andersson and Kokkotas made use of the universal behavior in the
QNMs of the fundamental fluid $f$-mode and the first polar
$w$-mode to estimate the radius and the mass of the star
\cite{Andersson1998}. Besides, the most probable EOS among all the
known models of stellar matters could also be identified from the
frequency of the leading $p$-mode (or axial $w$-mode) oscillation
\cite{Andersson1998,Ferrari}.

In a series of recent papers \cite{Tsui_1,Tsui_2}, we have studied
the physical origin of the universality in the pulsation
frequencies of the $f$-mode and $w$-mode stellar pulsations
discovered in Refs. \cite{Andersson1998} and \cite{Ferrari}, and
developed a scaled-coordinate logarithmic perturbative theory
(SCLPT) for neutron stars to locate axial $w$-mode oscillations.
The main objective of the current paper is then to work out a
feasible scheme to infer the internal structure of a neutron star
from its gravitational wave spectrum. Assuming that the
frequencies and the damping rates of a few $w$-mode oscillations
of a neutron star can be identified from gravitational-wave
observation, we iteratively invert the  SCLPT scheme to determine
the mass $M$, the radius $R$, the mass density $\rho(r)$ at radius
$r$, and the mass distribution function $m(r)$ (i.e. the total
mass inside radius $r$) of the star. In turn, accurate equation of
state of nuclear matter can also be obtained numerically from the
mass distribution function. While the idea underlying the
inversion scheme reported here is completely generic, we will
consider and use the spectrum of axial $w$-mode oscillations to
illustrate our method. As a remark, we note here that Lindblom has
proposed a scheme to determine, up to certain accuracy, EOS of the
nuclear matter of a neutron star from its mass and  radius
\citep{Lindblom_invert}. However, the EOS considered in Lindblom's
scheme is limited to barotropic type. In contrast, our method
directly exploit  the frequencies of several QNMs to determine the
radius,  the mass and other useful parameters of the star.
Moreover, the EOS obtained from our inversion scheme is free from
constraints. We will show in the following discussion that even
the exotic EOS of quark matter can be inferred from the
gravitational wave signals of a SQS.

The organization of our paper is as follows. In Sect.~II, we
review the universality existing in the QNM frequencies (including
the real and the imaginary parts) of neutron stars described by
different EOS and discuss how such behavior can be explained by
the Tolman VII model (TVIIM) of stars \cite{Tolman:1939jz}. In
Sect.~III we approximate the mass distribution of a realistic
neutron star by that of TVIIM plus some small correction terms to
be determined from the inversion scheme. Section IV outlines the
spirit of the inversion scheme while explicit formulas for the
scheme are derived in Sect.~V. In Sect.~VI we apply our method to
invert the EOSs of several realistic stars and numerical results
are shown there.  Section~VII studies the feasibility of our
method and then we conclude our paper in Sect.~VIII. The present
paper is the expanded version of a recent Letter \cite{Tsui_3}.
Unless otherwise stated, geometric units are used in this paper.

\section{Universality and TVIIM}
\begin{figure}
\includegraphics[angle=270,width=8.5cm]{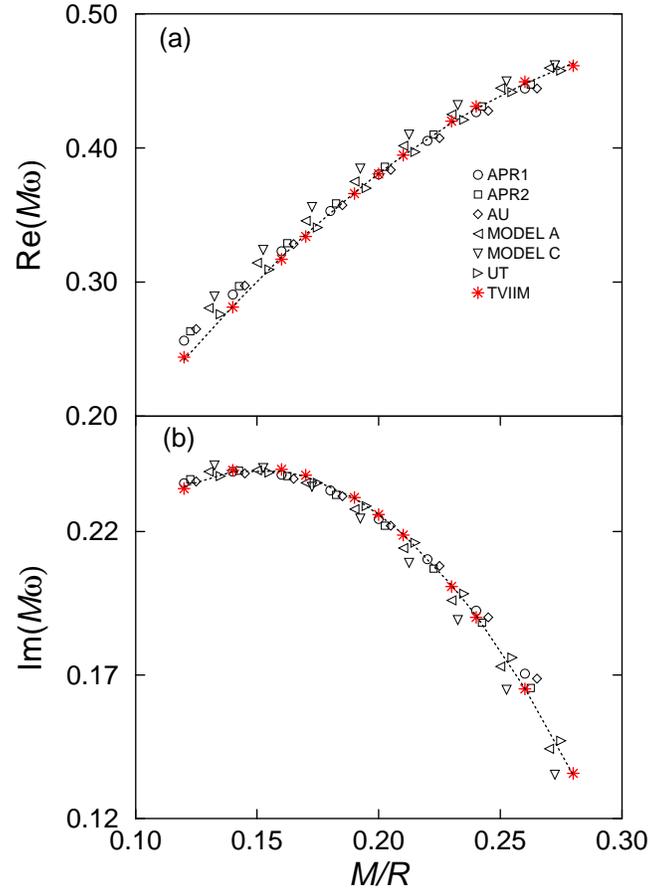}
\caption{The real and imaginary parts of $M\omega$ for the
least-damped axial $w$-mode of six realistic stars (unfilled
symbols) and TVIIM (stars) are shown as a function of $M/R$ in
panels (a) and (b) respectively. The dotted line represents the
best quadratic fit to those of TVIIM.} \label{f1}
\end{figure}

First of all, we review the universality in QNMs of gravitational
waves and state its physical origin, thus introducing the starting
point of our inversion scheme --- the TVIIM. It has recently been
observed that the QNM frequencies of the leading $w$-mode and
$f$-mode oscillations of non-rotating neutron stars approximately
satisfy the following scaling law \cite{Andersson1998,Ferrari}:
\begin{eqnarray}
M \omega = a
\left(\frac{M}{R}\right)^2+b\left(\frac{M}{R}\right)+c .
\label{TL}
\end{eqnarray}
Here  $a$, $b$, and $c$ are complex constants determined from
curve fitting. In Fig.~\ref{f1}, we plot $M\omega_{ r}$ and
$M\omega_{ i}$ against the compactness $\cc \equiv M/R$ for the
leading axial $w$-mode of non-rotating neutron stars characterized
by different EOSs, including models A and C proposed by
Pandharipande \cite{modelA,modelC}, models AU and UT proposed by
Wiringa, Fiks and Fabrocini \cite{AU}, models APR1 and APR2
proposed by Akmal, Pandharipande and Ravenhall \cite{APR}. One can
easily see that both $M\omega_{ r}$ and $M\omega_{ i}$ are, in
agreement with (\ref{TL}), well approximated by quadratic function
of $\cc$.

\begin{figure}
\includegraphics[angle=0,width=9cm]{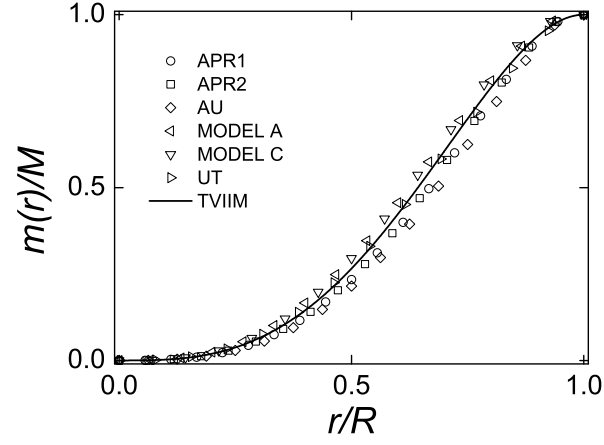}
\caption{The scaled mass distributions $m(r)/M$ of various
realistic neutron stars and TVIIM with a common compactness
$\cc=0.28$ are plotted against $r/R$.} \label{f3}
\end{figure}

The universality observed in  the spectra of realistic neutron
stars comes as a surprise because, as shown in Fig.~\ref{f2}, the
pressure $P$ at a given density $\rho$ varies greatly for the EOSs
considered there. To understand the physical origin underlying the
universality, we showed in Ref.~\cite{Tsui_1} that the scaled
complex eigenfrequencies $M\omega$ of axial $w$-mode, polar
$w$-mode and $f$-mode oscillations, to a good approximation,
depend only on the compactness $M/R$. In addition, we also
discovered that the TVIIM, which was first proposed by Tolman in
1939 \cite{Tolman:1939jz} and has a mass distribution function
given by:
\begin{equation}
m_c(r)= M \left[ \frac{5}{2}\left(\frac{r}{R}\right)^3
-\frac{3}{2}\left(\frac{r}{R}\right)^5 \right],
\end{equation}
can mimic the mass distribution of most neutron stars. As clearly
shown in Fig.~\ref{f3}, where the normalized mass function
$m(r)/M$ is plotted against the scaled radius $r/R$ for stars with
a compactness $M/R=0.28$ and described by different EOS as
mentioned above, the mass distribution function $m_c(r)$ is able
to approximate the mass distribution inside neutron stars with
different EOSs. Moreover, the scaled frequency $M\omega$ of QNMs
of TVIIM is close to those of realistic neutron stars (see
Fig.~\ref{f1}) and display similar scaling behavior mentioned
above. It is obvious that TVIIM does provide a good approximation
to stars with varying EOSs, and the universality shown in
Fig.~\ref{f1} can be captured by the best quadratic fit
 to the QNMs of TVIIM, with
  $a=-4.4-6.3{\rm
i}$, $b=3.1+1.9{\rm i}$, and $c=-0.072+0.098{\rm i} $. A distinct
feature of TVIIM  is that the scaled complex eigenfrequencies
$M\omega$ of its QNMs depend only on the compactness of the star
\cite{Tsui_1}. Hence, the observed universality in the QNMs
($w$-mode and $f$-mode) is expected to be closely related to TVIIM
and was indeed verified using SCLPT \cite{Tsui_2}.

The density distribution of TVIIM, given by
\begin{equation}\label{RealisticDrho}
    \rho(r)=\rho_0\left[1-\left(\frac{r}{R}\right)^2\right],
\end{equation}
with the central density $\rho_0=15 M/(8\pi  R^3)$, is so simple
that the metric coefficients $e^{\nu(r)}$ and $e^{\lambda(r)}$,
defined by the line element $\d s$:
\begin{equation}
\d s^2=-e^{\nu(r)}\d t^2+e^{\lambda(r)} \d r^2+r^2(\d
\theta^2+\sin^2\theta \d \varphi^2),
\end{equation}
can be obtained analytically:
\begin{eqnarray}
e^{-\lambda}&=&1-{\cal C} \xi^2(5-3\xi^2), \label{e_lam_eq}\\
e^{\nu}&=&(1-5{\cal C}/3)\cos^2\phi. \label{e_phi_eq}
\end{eqnarray}
Here $\xi=r/R$, and
\begin{eqnarray}
\phi&=&({w}_1-{w})/2+\phi_1, \\
{w}&=&\log\left[\xi^2-\frac{5}{6}+\sqrt{\frac{e^{-\lambda}}{3\cal{C}}}\right],\\
\phi_1&=&\phi(\xi^2=1)=\arctan\sqrt{\frac{\cal{C}}{3(1-2\cal{C})}}\,,\\
w_1&=&w(\xi^2=1)\,.
\end{eqnarray}
Besides, the pressure can also be found exactly from the
Tolman-Oppenheimer-Volkoff (TOV) equations
\cite{Oppenheimer:1939ne,Tolman:1939jz}:
\begin{eqnarray}
 P=\frac{1}{4\pi
R^2}\left[\sqrt{3\cc
e^{-\lambda}}\tan\phi-\frac{\cc}{2}(5-3\xi^2)\right],
\end{eqnarray}

By virtue of (\ref{RealisticDrho}), we have:
\begin{eqnarray}\label{}
\xi &=& \sqrt{1-\frac{\rho}{\rho_0}}, \\
e^{-\lambda}&=&1-{\cal C}
\left(1-\frac{\rho}{\rho_0}\right)\left(2+\frac{3\rho}{\rho_0}\right).
\end{eqnarray}
The EOS of TVIIM is consequently given by:
\begin{eqnarray}
 P=\frac{1}{4\pi
R^2}\left[\sqrt{3\cc -3{\cal C}^2
\left(1-\frac{\rho}{\rho_0}\right)\left(2+\frac{3\rho}{\rho_0}\right)}\tan\phi-\cc\left(1+\frac{3\rho}{2\rho_0}\right)\right],
\end{eqnarray}
with
\begin{eqnarray}
\phi&=&\frac{1}{2}\log\left[\frac{1}{6}+
\sqrt{\frac{1-2\cc}{3\cal{C}}}\right]+
\arctan\sqrt{\frac{\cal{C}}{3(1-2\cal{C})}} \nonumber \\
&&- \frac{1}{2}\log\left[\frac{1}{6}-\frac{\rho}{\rho_0}+\sqrt{
\frac{1}{3\cc}-\frac{1}{3}
\left(1-\frac{\rho}{\rho_0}\right)\left(2+\frac{3\rho}{\rho_0}\right)
}\right].
\end{eqnarray}
It is obvious that both $m_c(r)$ and the EOS for TVIIM depend on
$M$ and $R$ (or any two independent parameters derivable from
them). Thus, once these two parameters are known, the mass
distribution inside the star and the EOS of the stellar matter are
determined automatically. In the following discussion, we will
make use of this salient feature of TVIIM as the foundation of the
inversion scheme.
\section{Approximate Mass Distribution}\label{apprx mass}
\begin{figure}
\includegraphics[angle=0,width=9cm]{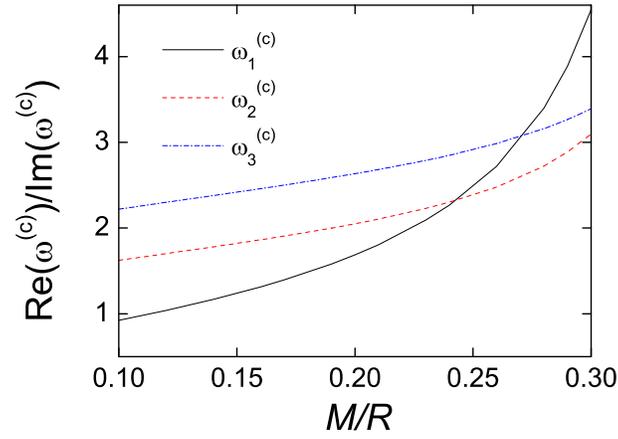}
\caption{${\rm Re} (\omega_{}^{(c)})/{\rm Im} (\omega_{}^{(c)})$
for the three leading modes of TVIIM star (i) $\omega_1^{(c)}$
(solid-line); (ii) $\omega_2^{(c)}$ (dashed-line) and (iii)
$\omega_3^{(c)}$ (dot-dashed-line) are plotted against the
compactness $M/R$.} \label{f4}
\end{figure}

Based on the the universality and the TVIIM discussed above, a
feasible scheme is proposed here to infer the internal structure
of a neutron star from its gravitational wave spectra. As the
first step of the scheme, we consider the frequency of the leading
(i.e. the least-damped)
 axial $w$-mode for TVIIM,
$\omega_1^{(c)}=\omega_{1r}^{(c)}-i\omega_{1i}^{(c)}$, and plot
the ratio $Q\equiv\omega_{1r}^{(c)}/\omega_{1i}^{(c)}$ against
with the compactness in Fig.~\ref{f4}. We see that
$\omega_{1r}^{(c)}/\omega_{1i}^{(c)}$ is in fact a monotonically
function of $M/R$. Hence, once $\omega_1^{(c)}$ is known, $M/R$
can be obtained from Fig.~\ref{f4} and in turn $M \omega_1^{(c)}$
 can be read from Fig.~\ref{f1}. It is then straightforward
 to find $M$ and $R$. Likewise, $M$ and $R$ can  be inferred from
 $\omega_2^{(c)}$ and $\omega_3^{(c)}$,
 the frequencies of the second and the third leading modes, as
 $\omega_{2r}^{(c)}/\omega_{2i}^{(c)}$ and $\omega_{3r}^{(c)}/\omega_{3i}^{(c)}$ are also monotonic
 functions of the compactness $\cc$ (see Fig.~\ref{f4}). However, the
 frequency $\omega_1^{(c)}$ is preferred in the subsequent
 discussion because it has the most sensitive dependence on $\cc$.

As TVIIM provides  a benchmark for other realistic stars, we
expect that the eigenfrequency of the leading axial $w$-mode
emitted from a realistic neuron star,
$\omega_1=\omega_{1r}-i\omega_{1i}$, is close to that of a TVIIM
star with the same mass $M$ and the same radius $R$. Therefore, we
could go through the procedure outlined in the last paragraph with
$\omega_1^{(c)}$ replaced by $\omega_1$ to obtain an estimate of
the mass and the radius of the star in consideration. This first
step in our scheme is analogous to the method proposed by
Andersson and Kokkotas \cite{Andersson1998}. However, we will
consider the frequencies of the non-leading modes and show that
they can lead to much improved estimates of the mass, the radius
and the EOS as well.

The mass distribution of a realistic neutron star is of course
different from that of a TVIIM star. However, the difference is
small. Considering the asymptotic behavior of the density around
$r=0$ (see, e.g. \cite{Chandrasekhar1}), we expand $m(r)$ as:
\begin{equation}
m(r,\{\mu_k\},M,R) \approx m_0(r,M,R)+\sum_{j=1}^{p-2}\mu_j
m_j(r,M,R),
\end{equation}
where
\begin{equation}
m_j(r,M,R)=\frac{15Mr}{2R^5}\left(\frac{r}{R}\right)^{2j}(r^2-R^2)^2,
\end{equation}
$p\geq2$, and $\mu_j$ ($j=1,2,\ldots,p-2$) are $p-2$ adjustable
parameters to provide the best fit to $m(r,\{\mu_k\},M,R)$. As the
mass $M$ and the radius $R$ of the neutron star are not yet
exactly known, they are also considered as free parameters. There
are consequently $p$ free parameters to be determined from the QNM
frequencies. Besides, as the zeroth order approximation to $m(r)$,
$m_0$ is well approximated by $m_c(r)$, which will be chosen to be
the initial guess for $m_0$.

Generally speaking, we can approximate the configuration of a
neutron star by choosing a set of $\mu_j$, $M$ and $R$ which
``best fit" the exact one. Before working out the optimization
process that determines these $p$ parameters, we briefly mention
the physical significance of $m_j(r,M,R)$. It is clear that
$m_j(r=R,M,R)=0$ for all $j$. Hence, $m_j$ does not affect the
total mass of the neutron star while modifying the internal mass
distribution. Furthermore, it can be shown that each $m_j$ has
only one maximum value and the position of such maximum gets
closer to the stellar surface as $j$ increases. This implies that
$m_j$ with small $j$ has significant contribution to the core of
the neutron star, while $m_j$ with large $j$ mainly affects the
the mass distribution near the stellar surface. In addition, the
central density of the neutron star depends only on $m_0$ and
$m_1$. As a consequence, as far as the high density regime of the
EOS  is concerned, $m_j$ with smaller $j$ is likely to be more
important. Lastly, the density distribution can be easily derived
from the mass distribution,
\begin{equation}
\rho(r,\mu_k,M,R)=\rho_0(r,M,R)+\sum_{j=1}^{p-2}\mu_j
\rho_j(r,M,R),
\end{equation}
where
\begin{equation}
\rho_0(r,M,R)=\frac{15M}{8\pi
R^3}\left[1-\left(\frac{r}{R}\right)^2\right],
\end{equation}
\begin{equation}
\rho_j(r,M,R)=\frac{15M}{8\pi R^7}\left(\frac{r}{R}\right)
^{2j-2}(r^2-R^2)\left[(5+2j)r^2-(1+2j)R^2\right].
\end{equation}

\section{The inversion Scheme}\label{InvertingScheme}
\subsection{Optimal model} Suppose that the mass and the
radius of a star are known with certain accuracy from the the
frequency of its leading mode, which are equal to $M_0$ and $R_0$,
respectively. The exact values of these quantities, $M=M_0+M_1$
and $R=R_0+R_1$, and the mass distribution inside the star can be
determined from the frequencies of QNMs of higher orders. Suppose
$n$ ($n>1$) QNM frequencies of a neutron star are known and we are
going to search for the optimal stellar model that yields the
minimal deviation from the measured QNM frequencies. As discussed
above, there are $p$ independent
 small parameters in the model, namely $\mu_j$
($j=1,2,\ldots,p-2$), $\mu_{p-1} \equiv M/R-M_0/R_0 \equiv
\cc^{(1)} $ , and $\mu_p \equiv  M_1$. If all of these free
parameters equal zero, the frequencies are denoted by
$\omega_q^{(0)}$.

Based on the first order result of SCLPT \cite{Tsui_2}, the
corresponding first order change in the $q$-th ($q=1,2,\ldots,n$)
QNM frequencies $\omega_q$ is
\begin{equation}\label{cqj}
\omega_q^{(1)} = \sum_{j=1}^{p}\mu_j\frac{\d\omega_q}{\d\mu_j}
=\sum_{j=1}^{p}\mu_j c_{qj},
\end{equation}
which is close to the exact change:
\begin{equation}
\Delta_{q}=\omega_q-\omega_{q}^{(0)}.
\end{equation}
Explicit expressions for $c_{qj}\equiv{\d\omega_q}/{\d\mu_j}$ will
be derived in the following section of this paper using SCLPT. By
minimizing the square-deviation
\begin{widetext}
\begin{eqnarray}\label{S}
    \chi^2
    &=& \sum^{n}_{q=1}|\Delta_q-\omega_q^{(1)}|^2 \nonumber\\
    &=& \sum^{n}_{q=1}\left(\Delta_q \bar{\Delta}_q- \Delta_q
     \sum^{p}_{j=1}
    \mu_{j}\bar{c}_{qj}
    -\bar{\Delta}_q\sum^{p}_{i=1} \mu_{i}c_{qi}+\sum^{p}_{i=1}
    \mu_{i}c_{qi}\sum^{p}_{j=1}
    \mu_{j}\bar{c}_{qj}\right),
\end{eqnarray}
\end{widetext}
where the overbar  indicates the complex conjugate of a variable,
one can obtain a set of $\mu_{j}$ and hence an optimal approximate
model for the observed neutron star. The necessary condition for
$\chi^2$ to achieve a minimum is obviously:
\begin{equation}
    \frac{\partial \chi^2}{\partial \mu_k}=0,
\end{equation}
for all $k$, leading to $n$ linear equations:
\begin{equation}\label{normaleqn}
    \sum^{n}_{q=1}\left( \Delta_q
    \bar{c}_{qk}
    +\bar{\Delta}_q c_{qk}\right)
    = \sum^{p}_{j=1} \mu_j \sum^{n}_{q=1}\left( c_{qk}\bar{c}_{qj}
    + \bar{c}_{qk}c_{qj} \right).
\end{equation}
This system of equations can be rewritten in the following matrix
representation:
\begin{equation}\label{matrix_A}
\vec b = {\bf A } {\vec \mu},
\end{equation}
where ${\bf A }$ is a $p \times p$ square matrix, $\vec \mu$ is
the column vector of $\{\mu_j \}$, and
\begin{eqnarray}
b_k &=& \sum^{n}_{q=1}\left( \Delta_q
    \bar{c}_{qk}
    +\bar{\Delta}_q c_{qk}\right), \\
A_{kj} &=& \sum^{n}_{q=1}\left( c_{qk}\bar{c}_{qj}
    + \bar{c}_{qk}c_{qj} \right),
\end{eqnarray}
for $k,j=1,2,3,\ldots,p$. Consequently, we can compute the best
$\mu_j$ by solving the above matrix equation.

\subsection{Iterative scheme}
With the $p$ parameters $\{\mu_j\}$ obtained through the scheme
sketched above, we can construct a new model star by incorporating
these mass corrections into $m_0(r)$ and compute a new set of QNM
frequencies for such a mass distribution. Again we can compare
these QNM frequencies with the observed QNM frequencies
$\omega_q$, and minimize the error $\chi^2$ by solving
(\ref{matrix_A}) to yield a new set of $\{\mu_j\}$. Through this
iterative scheme, one can generate  convergent sequences for the
mass, the radius and
 the mass distribution.
\subsection{Inversion of EOS}
The equilibrium configuration of a neutron star is given by  the
TOV equations \cite{Oppenheimer:1939ne,Tolman:1939jz}:
\begin{eqnarray}
\frac{\d m}{\d r} &=& 4\pi r^2\rho,\label{TOV1} \\
\frac{\d \nu}{\d r} &=& \frac{2m+8\pi r^3 P}{r(r-2m)},\label{TOV2} \\
\frac{\d P}{\d r} &=& -\frac{1}{2}(\rho+P)\frac{\d\nu}{\d
r}.\label{TOV3}
\end{eqnarray}
Here $\rho$ and $P$ are evaluated at a point $(r,\theta,\varphi)$.
Once $m(r)$ is known, the density $\rho(r)$ can be found from
(\ref{TOV1}). Combining (\ref{TOV2}) and (\ref{TOV3}) to yield a
first-order differential equation for $P(r)$:
\begin{eqnarray}
\frac{\d P}{\d r} = -(\rho+P)\frac{m+4\pi r^3
P}{r(r-2m)},\label{TOV2+3}
\end{eqnarray}
one can obtain $P(r)$  by solving (\ref{TOV2+3}) with the boundary
condition $P(r=R)=0$. Hence, the EOS can  be straightforwardly
inferred by comparing the two functions $\rho(r)$ and $P(r)$.

\section{SCLPT}
\subsection{Scaled Regge-Wheeler equation}
The starting point of the present inversion scheme is (\ref{cqj}),
which expresses the first-order shift in QNM frequencies as a
linear sum over various perturbations $\{\mu_j \}$. To evaluate
the coefficients $\{c_{qj} \}$ in (\ref{cqj}), we have to invoke
the theory of first-order SCLPT  to find how various perturbations
$\{\mu_j \}$ affect the QNM frequency  \cite{Tsui_2}.

First of all, we introduce the following scaled variables:
\begin{eqnarray}
\tr &=& \frac{r}{M},  \\
\tx &=& \frac{r_*}{M},  \\
\tm(\tr) &=& \frac{m(r)}{M},  \\
\tP(\tr) &=& M^2 P(r),  \\
\trho(\tr) &=& M^2 \rho(r),  \\
 \tilde{\nu}(\tr) &=& \nu(r),\\
 \tomega &=& M \omega,\\
\tilde{V}(\tilde{r}_*)&=&M^2 V(r_*).
\end{eqnarray}
Here the tortoise coordinate $r_{*}$  is related to the
circumferential radius $r$ by
\begin{equation}\label{r*_in}
r_{*}=\int_{0}^r \e^{(-\nu+\lambda)/2} \d r,
\end{equation}
and outside the star, the tortoise radial coordinate reduces to
\begin{equation}\label{r*_out}
r_{*}=r+2M \ln\Big(\frac{r}{2M}-1\Big)+C',
\end{equation}
where $C'$ is a constant that can be obtained by matching
(\ref{r*_in}) with (\ref{r*_out}) at $r=R$. Besides, we also let
$\tR \equiv R/M$ and $\tX \equiv R_*/M$, where $R_*=r_{*}(r=R)$ is
the tortoise radius of the star. In terms of these scaled
variables, the neutron star Regge-Wheeler (NSRW) equation
\cite{Chandrasekhar1} governing the evolution of axial $w$-mode
pulsation can be rewritten as \cite{Tsui_2}:
\begin{equation}\label{KG_sc}
\Big[\frac{\d^2}{\d\tilde{r}_*^2}+\tilde{\omega}^2-\tilde{V}(\tilde{r}_*)\Big]
\tilde{\psi}(\tilde{r}_*)=0,
\end{equation}
where for $\tr \le \tR $ the scaled Regge-Wheeler potential takes
the form:
\begin{equation}\label{sVin}
\tilde{V}_{*}(\tilde{r}_{})=\frac{e^{\tilde{\nu}}}{\tilde{r}^3}
[l(l+1)\tilde{r}+4\pi
\tilde{r}^3(\tilde{\rho}-\tilde{P})-6\tilde{m}(r)],
\end{equation}
and for $\tr>\tR$,
\begin{equation}\label{sVout}
\tV(\tx)=\left(1-\frac{2}{\tr}\right)\left[\frac{l(l+1)}{\tr^2}-
\frac{6}{\tr^3}\right].
\end{equation}

The scaled NSRW equation (\ref{KG_sc}) is obviously more amenable
to perturbative expansion since outside the star $\tV(\tr)$
depends only on $\tr$. Hence, changes in the stellar mass do not
directly affect the potential outside the star. Therefore,
perturbation arising from the change in the stellar mass, which
also leads to variations in mass density, pressure, and stellar
radius, becomes localized inside the star. In the subsequent
discussion, we will show that shifts in QNM frequencies can be
expressed in terms of integrals with finite domains of
integration.

\subsection{First order perturbation theory}
Consider an unperturbed neutron star which has a scaled
circumferential (tortoise) radius $\tilde{R}_{0}$
($\tilde{R}_{*0}$), a scaled potential
$\tilde{V}_0(\tilde{r}_{*})$, and  a scaled known mass
distribution function $\tilde{m}_0(r)$. Note that all physical
quantities are scaled with its own mass $M_0$. The unperturbed
star (e.g. TVIIM) will be used to provide the zeroth-order
approximation of a realistic star whose physical parameters are
not exactly known and to be determined from the inversion scheme.

Analogous to other conventional perturbation theories, we will
assume that the QNM wave function of the unperturbed star is
known. If now the star is perturbed to a new configuration which
is identical to that of the realistic star in consideration, the
new potential function is given by:
\begin{equation}\label{V_expansion}
\tV(\tx)=\tV^{(0)}(\tx)+\tV^{(1)}(\tx).
\end{equation}
As a consequence, the circumferential (tortoise) radius of the
perturbed star is also modified such that, $\tilde{R}=\tR_0+\tR_1$
($\tilde{R}_{*}=\tR_{*0}+\tR_{*1}$). However, outside the star,
$\tV$, as a function of $\tilde{r}$, remains unchanged.

 Consider the $q$-th QNM of
the unperturbed star, which has a scaled frequency
$\tomega_q^{(0)}$. From SCLPT we obtain the first-order shift in
the scaled frequency $\tomega \equiv M\omega$, which reads
\cite{Tsui_3}:
\begin{equation}\label{freq1}
\tomega_q^{(1)}=\frac{\langle\tpsi_q|\tU_1|\tpsi_q\rangle}
{2\tomega_q^{(0)}\langle\tpsi_q|\tpsi_q\rangle},
\end{equation}
with
\begin{equation}\label{U_1}
\langle\tpsi_q|\tU_1|\tpsi_q\rangle=\int_{0}^{\tilde{R}_{*0}}\d
\tx\tV_1(\tx)\tpsi_q^2(\tx)
+\tpsi_q^2(\tilde{R}_{*0})\left[\tR_{*1}\frac{\d
f_q^{(0)}}{\d\tx}-\tR_1\left(\frac{\partial
f_q^{(+)}(\tx,\tomega_q^{(0)})}{\partial\tr}\right)\right]_{\tx=\tilde{R}_{*0}}
,
\end{equation}
\begin{equation}\label{freq3}
\langle\tpsi_q|\tpsi_q\rangle=\int_0^{\tilde{R}_{*0}}\d
\tx\tpsi_q^2(\tx)+
\frac{\tpsi_q^2(\tilde{R}_{*0})}{2\tomega_q^{(0)}}
\left(\frac{\partial
f_q^{(+)}(\tilde{R}_{*0},\tomega)}{\partial\tomega}\right)_{\tomega=\tomega_q^{(0)}},
\end{equation}
and $\tpsi_q(\tx)$ being  the unperturbed wave function of the
$q$-th QNM mode. Besides, $f_q^{(0)}(\tx)$ is the logarithmic
derivative of the unperturbed wave function $\tpsi_q(\tx)$ that
satisfies both the regular boundary condition at $\tx=0$ and the
outgoing wave boundary condition at spatial infinity, while
$f_q^{(+)}(\tx,\tomega)$ is the logarithmic derivative of the
outgoing wave solution of frequency $\tomega$ to the scaled NSRW
equation \cite{Tsui_2}. From (\ref{freq1}), (\ref{U_1}) and
(\ref{freq3}) it is then obvious that one has to evaluate the
changes in the scaled potential, the scaled circumferential and
tortoise radii of the star in order to find $\tomega_q^{(1)}$.
\subsection{The perturbed star}
The perturbed star is in fact the realistic star considered in the
inversion scheme, whose  scaled mass distribution $\tm(\tr)$ and
density distribution $\trho(\tr)$ are expressed in terms of $\tr$
and its own exact compactness $\cc \equiv M/R$ as follows:
\begin{eqnarray}
\tm(\tr)&=&\frac{m(r)}{M}=\tm_0(\tr,\cc)+\sum_{j=1}^{p-2}\mu_j\tm_j(\tr,\cc), \\
\trho(\tr)&=&M^2\rho(r)=\trho_0(\tr,\cc)+\sum_{j=1}^{p-2}\mu_j\trho_j(\tr,\cc).
\end{eqnarray}
Here
\begin{widetext}
\begin{eqnarray}
\tm_0(\tr,\cc)&=&\frac{m_0(r,M,R)}{M}=\frac{1}{2}(5\cc^3\tr^3-3\cc^5\tr^5), \\
\tm_j(\tr,\cc)&=&\frac{m_j(r,M,R)}{M}=\frac{15}{2}(\cc\tr)^{1+2j}(1-\cc^2\tr^2)^2,\\
\trho_0(\tr,\cc)&=&M^2\rho_0(r,M,R)=\frac{15}{8\pi}\cc^3(1-\cc^2\tr^2),\\
\trho_j(\tr,\cc)&=&M^2\rho_j(r,M,R)=\frac{15}{8\pi}\cc^{2j+1}\tr^{2j-2}(\cc^2\tr^2-1)
\left[(5+2j)\cc^2\tr^2-(1+2j)\right].
\end{eqnarray}
\end{widetext}
In order to compare the perturbed and unperturbed stars, we
rewrite $\tm(\tr)$ and $\trho(\tr)$ as:
\begin{eqnarray}
\tm(\tr)&=&\tm^{(0)}(\tr)+\tm^{(1)}(\tr), \\
\trho(\tr)&=&\trho^{(0)}(\tr)+\trho^{(1)}(\tr),
\end{eqnarray}
with
\begin{eqnarray}
\tm^{(0)}(\tr)&=&\tm_0(\tr,\cc_0), \\
\trho^{(0)}(\tr)&=&\trho_0(\tr,\cc_0),\\
\tm^{(1)}(\tr)&=&\sum_{j=1}^{p-1}\mu_j\tm_j(\tr,\cc_0), \\
\trho^{(1)}(\tr)&=&\sum_{j=1}^{p-1}\mu_j\trho_j(\tr,\cc_0),
\end{eqnarray}
\begin{eqnarray}
\tm_{p-1}(\tr)\equiv\left(\frac{\partial \tm_0}{\partial
\cc}\right)_{\cc=\cc_0} &=&
\frac{15}{2}(\cc_0^2\tr^3-\cc_0^4\tr^5),\\
\trho_{p-1}(\tr)\equiv\left(\frac{\partial\trho_0}{\partial
\cc}\right)_{\cc=\cc_0}&=&\frac{15}{8\pi}\cc^2_0(3-5\cc_0^2\tr^2),
\end{eqnarray}
and $\mu_{p-1}\equiv\cc-\cc_0=\cc^{(1)}$ as mentioned previously.
$\tm^{(1)}(\tr)$ and $\tilde{\rho}^{(1)}(\tr)$ are considered as
the first order changes in the mass and density distributions.
Besides, it is straightforward to show that
\begin{equation}\label{}
\tilde{R}^{(1)}\equiv \frac{R}{M}-\frac{R_0}{M_0}
=-\frac{\cc^{(1)}}{\cc^{(0)}},
\end{equation}
\begin{equation}
\tX^{(1)}= \tR^{(1)}\exp\left[\frac{\tnu^{(0)}(\tR^{(0)})
-\tlam^{(0)}(\tR^{(0)})}{2}\right] +\int_0^{\tR^{(0)}}
\left[\frac{\tlam^{(1)}(\tr)-\tnu^{(1)}(\tr)}{2}\right]
\exp\left[\frac{\tlam^{(0)}(\tr)-\tnu^{(0)}(\tr)}{2}\right] d\tr.
\end{equation}

These first-order quantities, including $\tm^{(1)}, \trho^{(1)},
\tR^{(1)}, \tX^{(1)}$, are the inputs to SCLPT. Once
$\tomega^{(1)}_q$ is found from the first-order SCLPT, the first
order change in QNM frequency, $\omega^{(1)}$, can be obtained
from the following relation:
\begin{eqnarray}\label{}
\omega^{(1)}_q&=&\frac{\tomega^{(0)}_q+\tomega^{(1)}_q}{M^{(0)}+M^{(1)}}
-\frac{\tomega^{(0)}_q}{M^{(0)}}\nonumber \\
&=&\frac{\tomega^{(1)}_q}{M^{(0)}}-\frac{M^{(1)}\tomega^{(0)}_q}{[M^{(0)}]^2}
\end{eqnarray}

\subsection{Perturbed TOV equation}
The perturbation in $\tm^{(1)}(\tr)$ (or $\trho^{(1)}(\tr)$) of
course leads to variations in pressure and metric coefficients,
and in turn affects the potential $\tilde{V}$. So we need to
consider the scaled TOV equations \cite{Tsui_1,Tsui_2},
\begin{eqnarray*}
\frac{d\tm}{d\tr}&=&4\pi \tr^2\trho(\tr), \\
\frac{d\tnu}{d\tr}&=&\frac{2\tm(\tr)+8\pi \tr^3 \tP(\tr)}{\tr(\tr-2\tm(\tr))}, \\
\frac{d
\tP}{d\tr}&=&-\frac{1}{2}(\trho(\tr)+\tP(\tr))\frac{d\tnu}{d\tr},
\end{eqnarray*}
with $\tnu=\nu$. Denoting the first order changes in $\tP$ and
$\tnu$ by $\tP^{(1)}$ and $\tnu^{(1)}$, we can show from the
scaled TOV equations that:
\begin{widetext}
\begin{eqnarray}\label{pTOV}
\frac{d\tnu^{(1)}}{d\tr} &=&\frac{1}{(\tr-2\tm^{(0)})^2}\left[
(2+16\pi \tr^2\tP^{(0)})\tm^{(1)}+8\pi \tr^2(\tr-2\tm^{(0)})
\tP^{(1)}
\right], \\
\frac{d \tP^{(1)}}{d\tr}&=&-\frac{1}{2}\left[\left(\trho^{(1)}
+\tP^{(1)}\right)\frac{d\tnu^{(0)}}{d\tr}
+(\trho^{(0)}+\tP^{(0)})\frac{d\tnu^{(1)}}{d\tr}\right].
\end{eqnarray}
\end{widetext}
Therefore, both $\tP^{(1)}$ and $\tnu^{(1)}$  can be evaluated by
numerically solving this pair of differential equations. Besides,
from
\begin{equation}
e^{-\tlam(\tr)}=1-\frac{2\tm(\tr)}{\tr},
\end{equation}
we have
\begin{equation}
\tlam^{(1)}=\frac{2\tm^{(1)}\exp(\tlam^{(0)})}{\tr} .
\end{equation}

\subsection{Perturbed potential}
Upon the changes in mass and density functions, the scaled
$\tV_{rw}$ varies in accordance with (\ref{V_expansion}). However,
$\tV$ is an explicit function of $\tr$, instead of $\tx$. In order
to express $\tV_{}$  as a function of $\tx$ such that our
perturbation scheme is applicable, we have to  regard $\tr$ as a
function of $\tx$ and in turn obtain the implicit function
$\tV(\tr(\tx))$. In the absence of perturbation,
$\tr=\tr^{(0)}(\tx)$ is determined from the defining equation of
the tortoise coordinate:
\begin{equation}
\tx=\int_0^{\tr^{(0)}(\tx)}
\exp\{[-\tnu^{(0)}(\tr)+\tlam^{(0)}(\tr)]/2\}d\tr.
\end{equation}
With the introduction of $\tm^{(1)}$ and $\tilde{\rho}^{(1)}$, the
scaled circumferential radius is modified such that
$\tr=\tr^{(0)}(\tx)+\tr^{(1)}(\tx)$, where $\tr^{(1)}(\tx)$ is
defined via the following equation:
\begin{equation}
\tx=\int_0^{\tr^{(0)}(\tx)+\tr^{(1)}(\tx)}
\exp\{[\tlam^{(0)}(\tr)+\tlam^{(1)}(\tr)-\tnu^{(0)}(\tr)-\tnu^{(1)}(\tr)]/2\}d\tr.
\end{equation}
Neglecting second order terms, we can find $\tr^{(1)}$ in terms of
$\tx$:
\begin{equation}
\tr^{(1)}(\tx)= \exp\left[\frac{\tnu^{(0)}(\tR^{(0)})
-\tlam^{(0)}(\tR^{(0)})}{2}\right] \int_0^{\tr^{(0)}(\tx)}
\left[\frac{\tnu^{(1)}(\tr)-\tlam^{(1)}(\tr)}{2}\right]
\exp\left[\frac{\tlam^{(0)}(\tr)-\tnu^{(0)}(\tr)}{2}\right] d\tr.
\end{equation}

By virtue of (\ref{V_expansion}), we have
\begin{eqnarray}\label{V_expansion2}
\tV^{(1)}(\tx)&=&\tV(\tr^{(1)}+\tr^{(0)})-\tV^{(0)}(\tr^{(0)})\nonumber \\
&=&\left[\tV(\tr^{(0)})-\tV^{(0)}(\tr^{(0)})\right]+\tr^{(1)}\frac{d\tV^{(0)}}{d\tr^{(0)}},
\end{eqnarray}
and from (\ref{sVin}) explicit expressions for
 the terms in the right hand side of (\ref{V_expansion2}) can be
 found:
\begin{widetext}
\begin{eqnarray}
\frac{d\tV^{(0)}}{d\tr^{(0)}}&=&
\tV^{(0)}\frac{\partial\tnu^{(0)}}{\partial
\tr^{(0)}}+e^{\tnu^{(0)}} \left\{-\frac{2l(l+1)}{(\tr^{(0)})^3}
+4\pi\left(\frac{d\trho^{(0)}}{d \tr^{(0)}} -\frac{d \tP^{(0)}}{d
\tr^{(0)}}\right)-\frac{6[4\pi (\tr^{(0)})^3
\trho^{(0)}-3\tm^{(0)}]}{(\tr^{(0)})^4}\right\},
\end{eqnarray}
\begin{eqnarray}
 \tV(\tr^{(0)})-\tV^{(0)}(\tr^{(0)})&=&
\tnu^{(1)}\tV^{(0)}(\tr^{(0)})+e^{\tnu^{(0)}}
\left[4\pi\left(\trho^{(1)}
-\tP^{(1)}\right)-\frac{6\tm^{(1)}}{(\tr^{(0)})^3}\right].
\end{eqnarray}
\end{widetext}
The results obtained above together with the first order SCLPT
lead explicit expressions for $c_{qj}$ in (\ref{cqj}).

\section{Numerical results}
\begin{figure}
\includegraphics[angle=270,width=8.5cm]{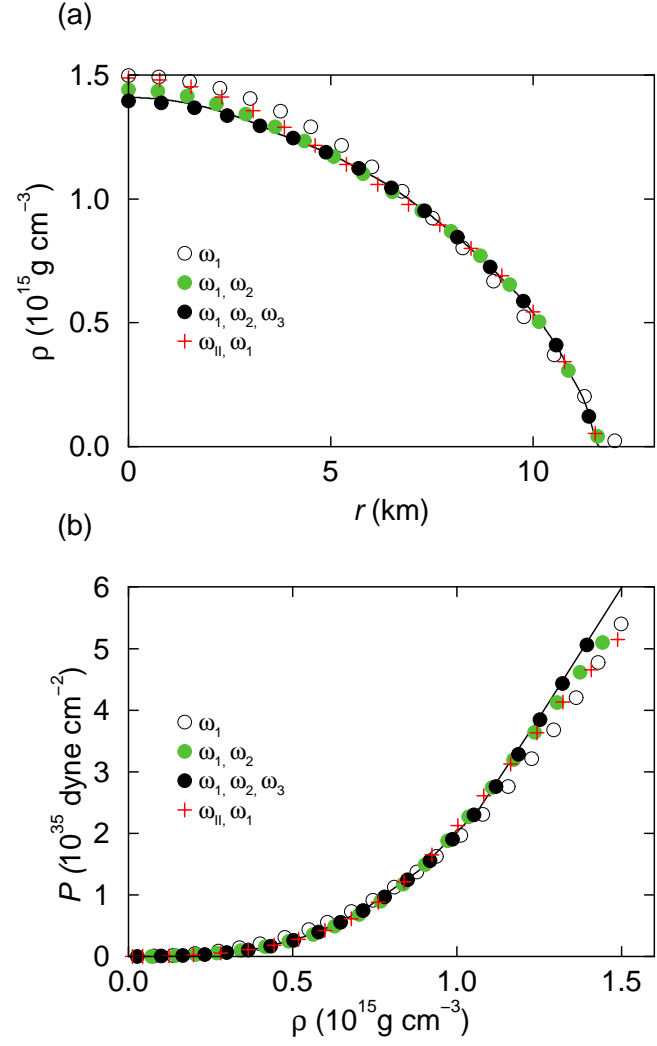}
\caption{Panels (a) and (b) depict $\rho(r)$ and $P(\rho)$
respectively for an APR1 star with $\cc=0.28$. The solid line is
the theoretical value, while the unfilled/grey/dark circles
represent the results obtained from inversion scheme using
one/two/three leading axial $w$-modes. The result obtained from
inversion scheme using $\omega_1$ and the frequency of a $w_{\rm
II}$-mode is shown by the crosses.} \label{f5}
\end{figure}

\begin{figure}
\includegraphics[angle=270,width=8.5cm]{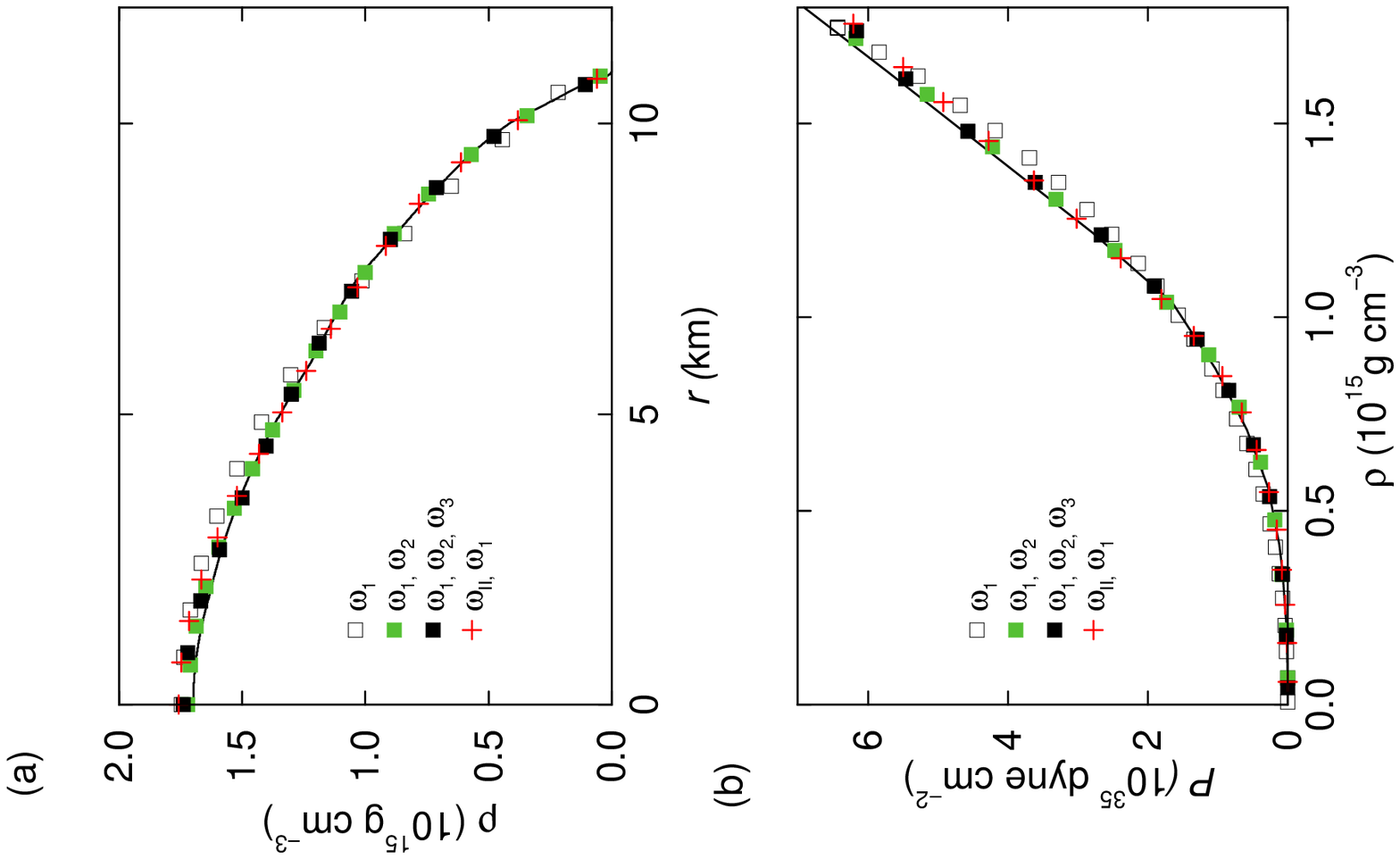}
\caption{Same as Fig.~\ref{f5}, except that an APR2 star is
considered instead.} \label{f6}
\end{figure}

\begin{figure}
\includegraphics[angle=270,width=8.5cm]{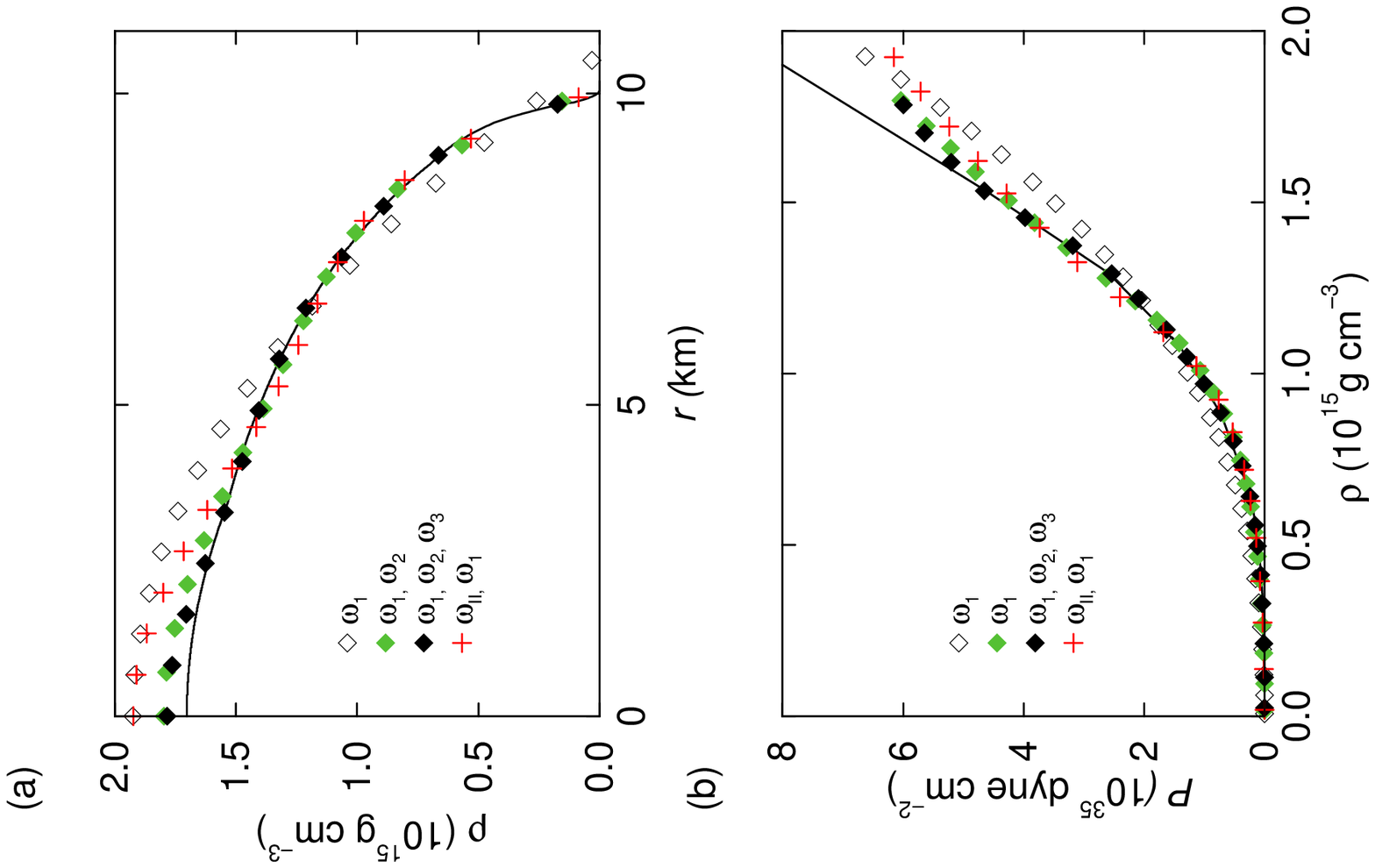}
\caption{Same as Fig.~\ref{f5}, except that an AU star is
considered instead.} \label{f7}
\end{figure}

\begin{figure}
\includegraphics[angle=270,width=8.5cm]{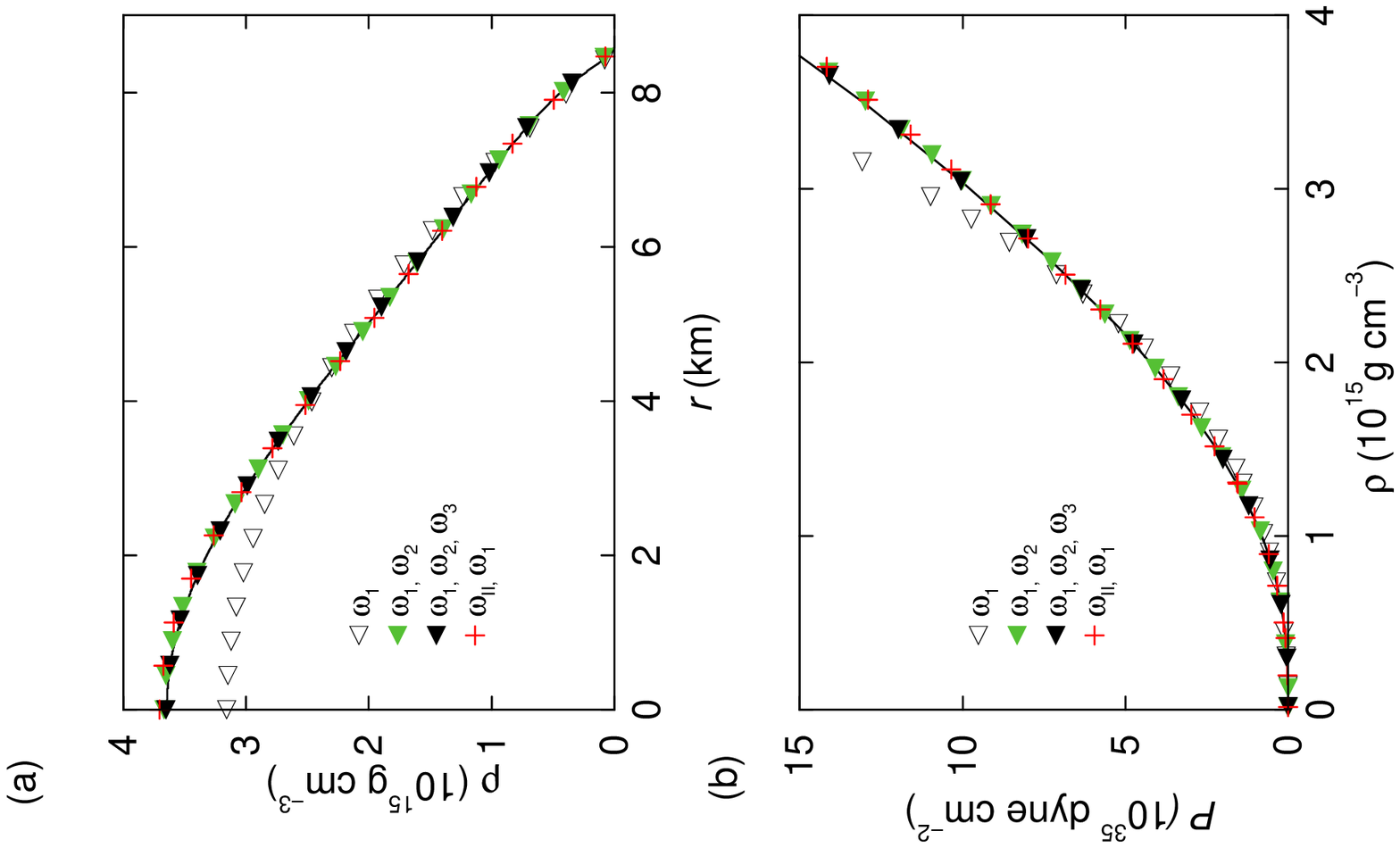}
\caption{Same as Fig.~\ref{f5}, except that a model A star is
considered instead.} \label{f8}
\end{figure}

\begin{figure}
\includegraphics[angle=270,width=8.5cm]{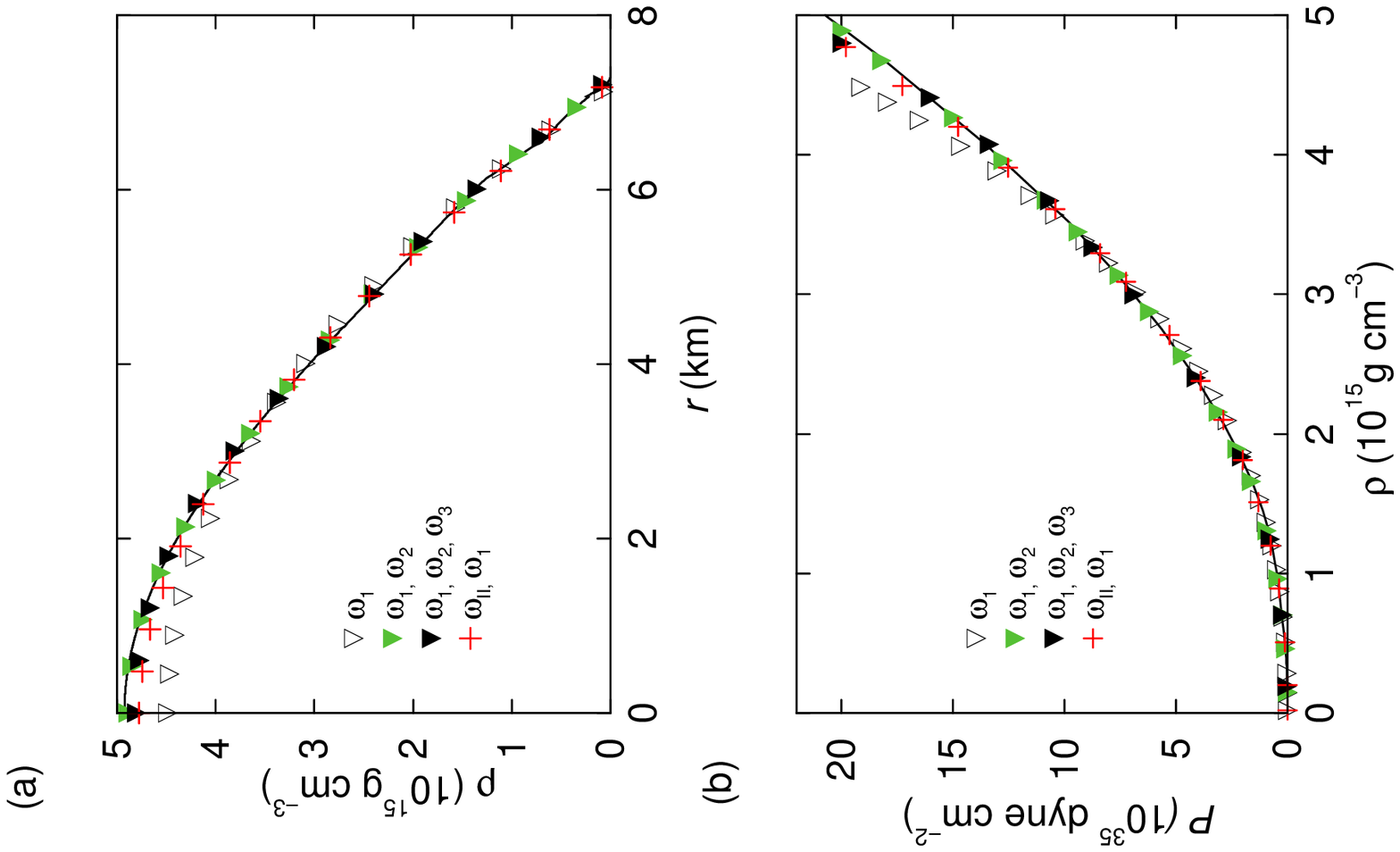}
\caption{Same as Fig.~\ref{f5}, except that a model C star is
considered instead.} \label{f9}
\end{figure}

\begin{figure}
\includegraphics[angle=270,width=8.5cm]{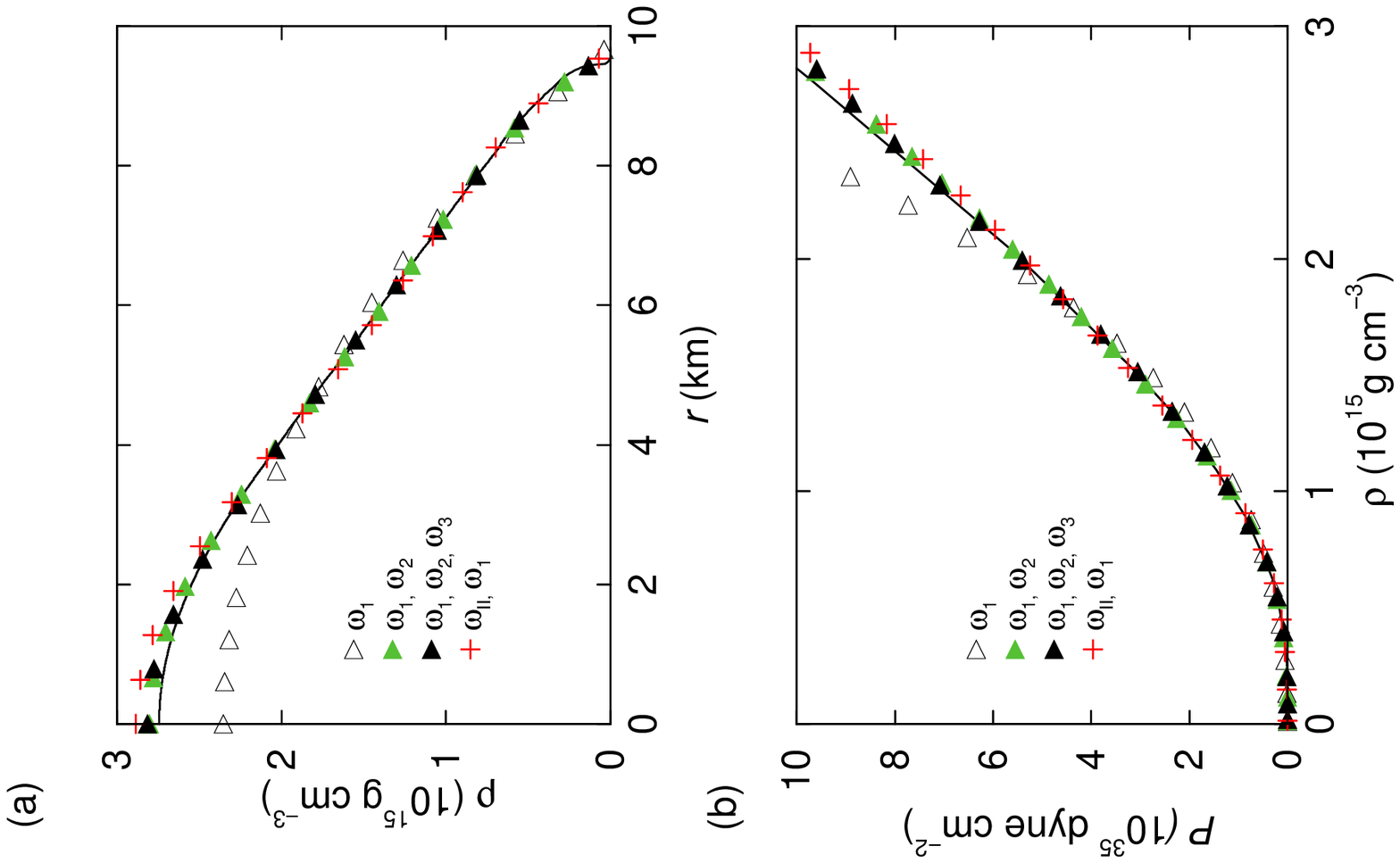}
\caption{Same as Fig.~\ref{f5}, except that a UT star is
considered instead.} \label{f10}
\end{figure}

In Figs.~\ref{f5} - \ref{f10}, we show numerical results obtained
from the inversion scheme developed in the present paper for six
different realistic neutron stars with a common compactness of
$0.28$. To gauge the accuracy of the scheme, we have used four
different combinations of QNM frequencies: (i) $\omega_1$; (ii)
$\omega_1$ and $\omega_2$; (iii) $\omega_1$, $\omega_2$ and
$\omega_3$; and (iv) $\omega_1$ and $\omega_{\rm II}$. Here
$\omega_1, \omega_2, \omega_3, \ldots $ are ordered in increasing
frequencies, and $\omega_{\rm II}$ is the frequency of a $w_{\rm
II}$ mode \cite{Leins}. In all cases the number of parameters $p$
was chosen to be twice the number of QNMs (i.e. $2n$). It is
clearly seen from these figures that (i) the results obtained from
using only the frequency of the least damped mode, $\omega_1$
(empty symbols), can nicely reproduce the EOS except at the high
density regime; (ii) the results obtained from a combination of
$\omega_1$ and $\omega_2$ (grey symbols) are readily improved;
(iii) using all three frequencies $\omega_1$, $\omega_2$ and
$\omega_3$ (dark symbols) indeed yields a perfect match with the
exact values; and (iv) the results obtained from the combination
of $\omega_1$ and $\omega_{\rm II}$ (crosses) are slightly less
accurate than those from the combination of $\omega_1$ and
$\omega_{2}$.

In Fig.~\ref{f2} we compare these six EOSs and demonstrate that it
suffices to use only two frequencies ($\omega_1$ and $\omega_2$)
to accurately reproduce and distinguish these EOSs. In Table~I we
tabulate the exact values of $M$ and $R$ and those obtained from
our scheme using one or two QNMs for stars constructed with these
EOSs. The result is truly encouraging, especially for the mass,
which is very close to the exact value in all cases. Despite that
the computed radius might deviate from the exact one by a few
percent in some cases,  the result is in fact very accurate
taking into account of the  fact that the mass density near the
surface of realistic neutron stars is usually very low. Hence, we
conclude that the frequencies of two leading $w$-mode oscillations
can readily lead to accurate determination of $M$, $R$ and EOS.

\begin{figure}
\includegraphics[angle=0,width=10cm,height=14.5cm]{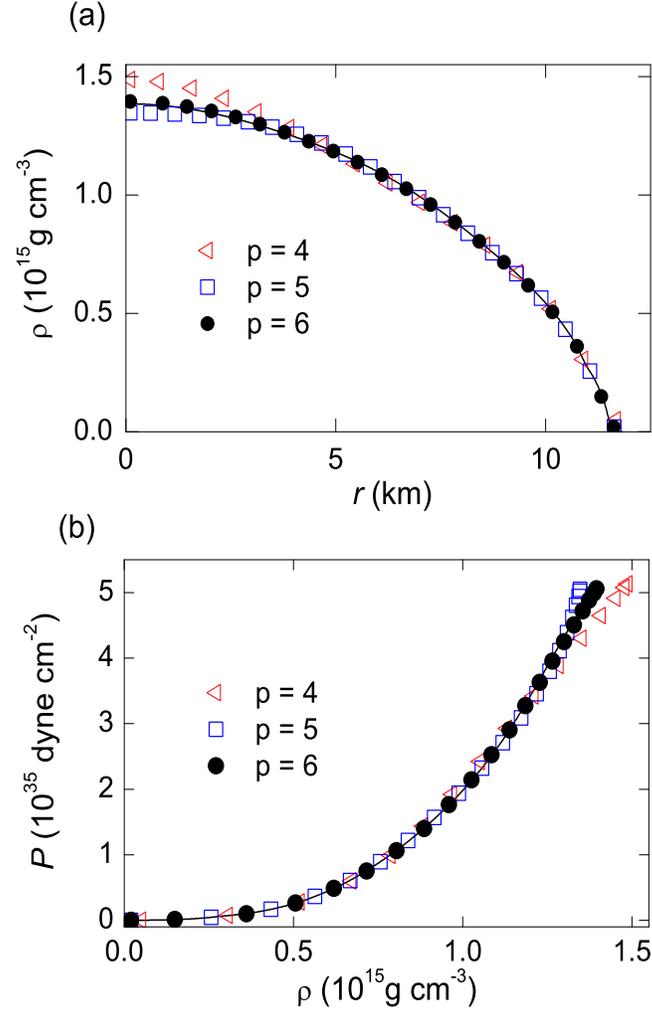}
\caption{Panels (a) and (b) depict $\rho(r)$ and $P(\rho)$
respectively for an APR1 star with $\cc=0.28$. The solid line is
the theoretical value, while the left triangles/squares/dark
circles represent the results obtained from inversion scheme using
four/five/six parameters and three leading axial $w$-modes.}
\label{f11}
\end{figure}

In each inversion process the number of parameters $p$ is less
than or equal to twice the number of QNMs used (i.e. $2 n$). This
can be easily understood as each QNM frequency carries two data,
namely the pulsation and decay rates. As a result, at most $2n$
parameters can be determined from the inversion process. On the
other hand,  as shown in Fig.~\ref{f11}, where the inversion
scheme is applied to an APR1 star with three known QNMs, the case
$p=6$ obviously yields the best result. Therefore, for a fixed
number of QNMs, the choice $p=2n$ seems to be the optimal one and
this remark is supported by our numerical results.

The stability of the inversion process has been investigated by
introducing artificial noises into the real and imaginary parts of
QNM frequencies. In each inversion process the radius of the star
in consideration, which was also an unknown to be determined from
the process,  was divided into a fixed number of equal partitions.
The values of the radius, density, and pressure at each of these
small partitions were in turn obtained from the inversion scheme.
After gathering the data, we averaged them over a number of
simulations in which random noises were superimposed on the values
of QNM frequencies, and evaluated the respective error bars.
Fig.~\ref{f12} shows the result for an APR1 star with two
approximately known QNM frequencies, $\omega_1$ and $\omega_2$,
and four parameters were used in the inversion. The random noise
introduced was within $\pm 5\%$ of the exact value. The grey
circles (with error bars) denote the average values of 12
simulations, which are fairly accurate. This strongly suggests
that the inversion scheme proposed here is still feasible even if
gravitational signals are blurred by experimental noises. However,
there was a need for caution to be exercised in inverting
noise-polluted QNM frequencies. In some cases the result obtained
might explicitly violate the thermodynamic stability condition
${\rm d}\rho/{\rm d}p$. We have in fact discarded such ill-defined
data in the inversion scheme mentioned above.

\begin{figure}
\includegraphics[angle=270,width=7.5cm]{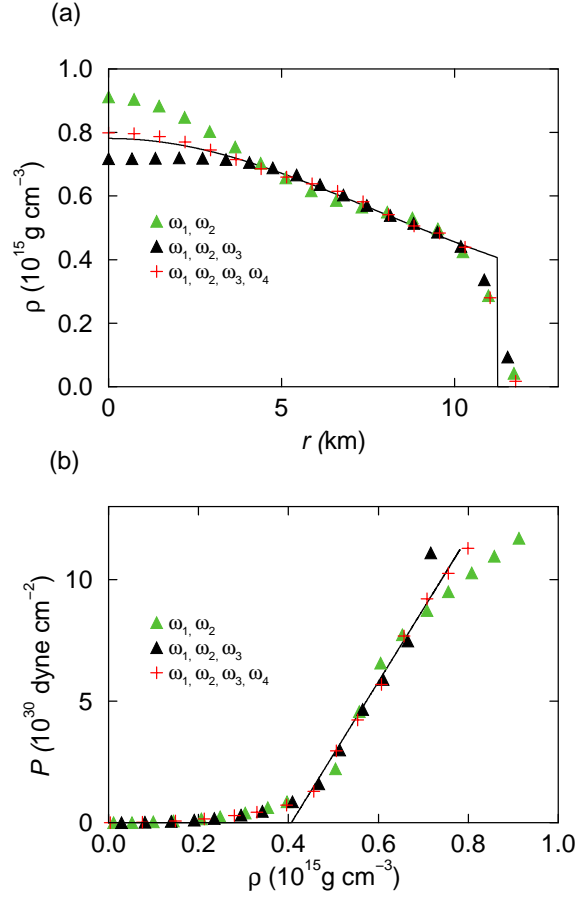}
\caption{Panels (a) and (b) depict $\rho(r)$ and $P(\rho)$
respectively for a SQS star with $\cc=0.2$. The solid line is the
theoretical value, while the grey triangles/dark triangles/crosses
respectively represent the results obtained from inversion scheme
using two, three and four leading axial $w$-modes.  } \label{f13}
\end{figure}

Lastly, to further demonstrate the robustness of the inversion
scheme, we have also applied our scheme to infer the EOS of a SQS
\cite{AFO}. The EOS of quark matter is given by the MIT bag model
\citep{MITBM}:
\begin{equation}\label{}
P=(\rho-4B)/3,
\end{equation} where the value of the bag
constant $B$ is  $57 \,{\rm MeV\, fm}^{-3}$. The crust of the SQS
star, a thin layer of normal matter obeying the
Baym-Pethick-Sutherland EOS \citep{ComStar}, is supported by
Coulomb forces above the quark matter core \cite{AFO}. The EOS for
quark matter is rather exotic and the density profile of such a
star is even discontinuous across the inner boundary of the crust.
However, as shown in Fig.~\ref{f13}, the inversion scheme still
works nicely as long as four QNMs are used. It is then obvious
that the current inversion scheme outperform the one proposed by
Lindblom, which is limited to barotropic type EOS
\citep{Lindblom_invert}.
\section{Detection of QNMs}
In the above discussion we show how the internal structure of a
neutron star can be systematically determined from  several of its
QNM frequencies.  Like all other studies in the area of
gravitational-wave asteroseismology (see e.g.
\citep{Andersson1996,Andersson1998,Ferrari,Kokkotas_2001,super_prl,
Kojima_02,Ferrari2003MNRAS,Sotani_03,Ferrari2004}), our method
depends crucially on reliable data of gravitational wave signals,
which have not yet been available. Therefore, the realization of
the present scheme hinges on the successful operation of various
existing gravitational-wave detectors (such as LIGO, VIRGO,
GEO600), and also the development in the design and construction
of new gravitational-wave detectors in the future. In the
following discussion, we consider the possibility of detecting
QNMs of neutron stars based on analysis outlined in
Ref.~\cite{Kokkotas_2001}.

\begin{figure}
\includegraphics[angle=0,width=9cm]{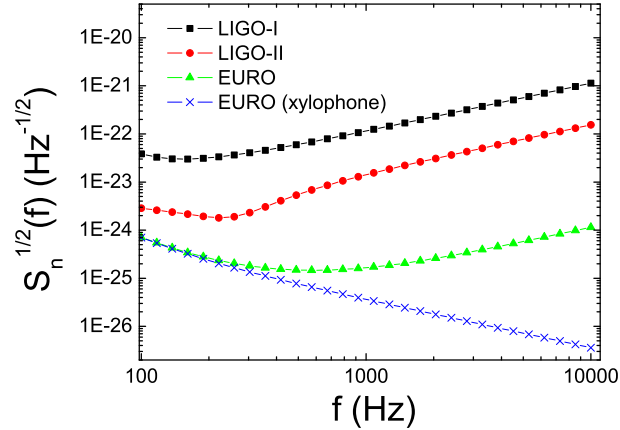}
\caption{The noise amplitude $S_n^{1/2}(f)\,({\rm Hz}^{-1/2})$ is
plotted against $f\,({\rm Hz})$ for gravitational-wave detectors
LIGO-I, LIGO-II, EURO and EURO (xylophone). The curves are based
on the interpolation formulas available at the EURO homepage
http://www.astro.cf.ac.uk/geo/euro.} \label{f14}
\end{figure}

The detection of a QNM signal, an exponentially decaying
sinusoidal wave with frequency $f=\omega_r/(2\pi)$ and decay time
$\tau=1/\omega_i$, depends on the distance $r$ between the emitter
(the neutron star) and the observer (the earth), the energy
$E_{\rm gw}$ channelled  in such a QNM, and of course the
sensitivity of the gravitational-wave detector, which is measured
in terms of its noise power spectral density $S_n(f)$ (see
Fig.~\ref{f14}). A strong signal is associated with a large
signal-to-noise ratio given by \cite{Kokkotas_2001}:
\begin{equation}\label{S/N}
\left( \frac{S}{N}\right)^2=\frac{4Q^2}{1+4Q^2}\frac{A^2
\tau}{2S_n},
\end{equation}
where the quality factor $Q=\pi f \tau$, and the amplitude of such
a gravitational wave signal \cite{Kokkotas_2001}:
\begin{equation}\label{amp}
A \sim 2.4 \times 10^{-20}\left(\frac{E_{\rm gw}}{10^{-6}{\rm
M}_\odot c^2}\right)^{1/2}\left(\frac{10\, {\rm kpc}}{r}\right)
\left(\frac{1\, {\rm kHz}}{f}\right)\left(\frac{1\, {\rm
ms}}{\tau}\right)^{1/2}.
\end{equation}
The fractional statistical errors  in the determination of the
frequency and the decay time, respectively denoted by
${\sigma_f}/{f}$ and ${\sigma_\tau}/{\tau}$, are estimated to be
\cite{Kokkotas_2001}:
\begin{equation}\label{error_f}
\frac{\sigma_f}{f} \simeq 0.0042
P^{-1}\sqrt{\frac{1-2Q^2+8Q^4}{4Q^4}} \left(\frac{\tau}{1\, {\rm
ms}}\right)^{-1},
\end{equation}
\begin{equation}\label{error_T}
\frac{\sigma_\tau}{\tau} \simeq 0.013
P^{-1}\sqrt{\frac{10+8Q^2}{Q^2}} \left(\frac{f}{1\, {\rm
kHz}}\right),
\end{equation}
with
\begin{equation}\label{}
P^{-1}=\left(\frac{E_{\rm gw}}{10^{-6}{\rm M}_\odot
c^2}\right)^{-1/2}\left(\frac{r}{10\, {\rm kpc}}\right)
\left(\frac{S_n^{1/2}}{10^{-23}\, {\rm Hz}^{-1/2}}\right).
\end{equation}

Both LIGO and VIRGO have a broad-band sensitivity in the
hundred-Hertz range where $S_n^{1/2} \sim 10^{-23}\, {\rm
Hz}^{-1/2}$ (see Fig.~\ref{f14}). It is then obvious that these
first generation detectors could not accurately locate $w$-mode
QNMs of neutron stars in our galaxy, which are characterized by
the following typical values: $f \sim 10 \,{\rm kHz}$, $ \tau \sim
0.02 \,{\rm ms}$, $E_{\rm gw} \sim {10^{-6}{\rm M}_\odot c^2}$ and
${r} \sim 10\, {\rm kpc}$ \cite{Kokkotas_2001}. In fact, it is
generally believed that the sensitivities of the first generation
detectors are not sufficient to carry out quantitative
measurements.

LIGO-II, a second generation gravitational-wave detector,  is
aimed at operation in the hundred-Hertz range near the standard
quantum limit (SQL) and has a lower noise level given
approximately by \cite{Cardiff}:
\begin{equation}\label{LIGOII}
\frac{S_n(f)}{\rm
Hz^{-1}}=10^{-49}\left[x^{-4.14}-5x^{-2}+\frac{(111)(2-2x^2+x^4)}{2+x^2}
\right],
\end{equation}
with $x=f/(215 \,{\rm Hz})$. In spite of the improved sensitivity,
as shown in Table~\ref{t2}, where $E_{\rm gw} = {10^{-6}{\rm
M}_\odot c^2}$ and ${r} = 10\, {\rm kpc}$ are assumed, LIGO-II is
not yet sensitive enough to locate $w$-mode QNMs of neutron stars
in our own galaxy. On the other hand, it is worthwhile to note
that SQL can in fact be overcome by utilizing quantum
non-demolition (QND) devices \cite{Bra}. The design of LIGO-III, a
third generation gravitational-wave detector, is currently under
intense discussion. Various strategies that can beat the
gravitational-wave SQL by an arbitrarily large amount, over an
arbitrarily wide range of frequencies have been proposed and
analyzed \cite{GWQND_T,GWQND}.

In the present paper we consider specifically another third
generation gravitational-wave detector, EURO (European
Gravitational Wave Observatory), proposed by a group of European
scientists in 2000, aims to make quantitative surveys of
gravitational-wave signals over 4 decades of frequency, namely
from 1 Hz to tens of kHz \cite{super_prl,Cardiff}. As stated in
the proposal \cite{Cardiff}, {\it ``EURO can do asteroseismology
on neutron stars by resolving the frequencies of the dominant
normal modes of vibration"}. In the proposal two possible
configurations of EURO were discussed. In the first configuration,
EURO, the sensitivity in the high frequency range is still limited
by the photon shot noises, with a noise level given by
\cite{Cardiff}:
\begin{equation}\label{E1}
\frac{S_n(f)}{\rm
Hz^{-1}}=10^{-50}\left[\frac{3.6\times10^{9}}{f^4}+ \frac{1.3
\times 10^5}{f^2}+1.3 \times 10^{-3}f_k
\left(1+\frac{f^2}{f_k^2}\right)\right],
\end{equation}
where the knee-frequency $f_k=1000\,{\rm Hz}$. As shown in
Fig.~\ref{f14}, the noise level of this configuration in the kHz
range is two orders less  than that of LIGO-II. With this detector
$w$-mode gravitational wave signals emitted  from neutron stars
within our galaxy  can be detected with error bound of a few
percent. Therefore, the inversion scheme outlined above will
become viable once EURO operates as proposed.

In the second configuration, referred to as EURO (xylophone) in
the present paper, the shot noise limit in the high-frequency
regime is beaten by parallel operation of several narrow-banded
cryogenic interferometers comprising a ``gravitational-wave
xylophone" \cite{super_prl,Cardiff}. The noise level of such a
device is approximately given by \cite{Cardiff}:
\begin{equation}\label{EX}
\frac{S_n(f)}{\rm
Hz^{-1}}=10^{-50}\left[\frac{3.6\times10^{9}}{f^4}+ \frac{1.3
\times 10^5}{f^2}\right],
\end{equation}
which can be obtained by removing the last term (i.e. the
shot-noise term) in (\ref{E1}). As can be inferred from
Fig.~\ref{f14}, Table~\ref{t2}, and also the following scaling
relations \cite{Kokkotas_2001}:
\begin{eqnarray}\label{}
  \left(\frac{S}{N}\right)^2 &\propto& \frac{E_{\rm gw}}{r^2},\\
  \left(\frac{\sigma_f}{f}\right) &\propto& \frac{r}{\sqrt{E_{\rm
  gw}}},\\
\left(\frac{\sigma_\tau}{\tau}\right) &\propto&
\frac{r}{\sqrt{E_{\rm
  gw}}},
\end{eqnarray}
 the advent of the detector EURO (xylophone) will
enable observation of gravitational waves from extra-galactic
neutron stars and black holes. We therefore expect that the
inversion scheme developed in the present paper can reveal
fruitful information about the internal structure of neutron stars
from the data gathered by these third generation
gravitational-wave detectors.
\section{Conclusion}
\begin{figure}
\includegraphics[angle=270,width=7.5cm]{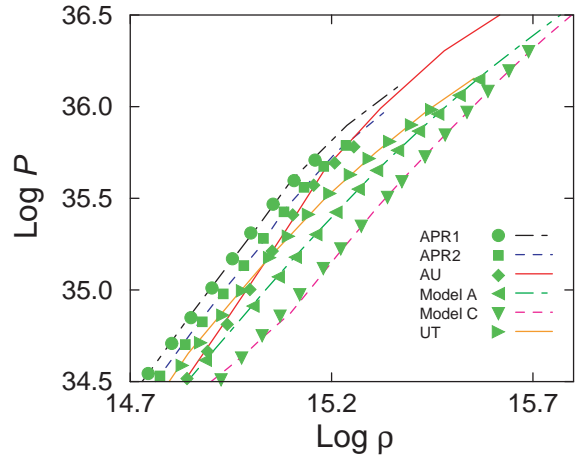}
\caption{Six different EOSs are shown by  lines of different
formats. Grey symbols are the corresponding values obtained from
inversion scheme using $\omega_1$ and $\omega_2$ for stars with
$\cc=0.28$. Here $P$ and $\rho$ are in cgs units.} \label{f2}
\end{figure}

Despite the mass of some neutron star binaries can be inferred
from their orbital periods, as yet there is no generic method to
determine their radii except for a few cases where data of
gravitational red-shift are available \cite{Ozel,Cottam}. In the
present paper we have proposed a robust inversion scheme to
determine the mass, the radius, the mass distribution and the EOS
of a neutron star from its gravitational wave spectra, which could
be obtained by the matched filtering method (see, e.g.
\cite{Kokkotas_2001}). We expect that our scheme could operate in
conjunction with gravitational-wave detectors available in one or
two decades, e.g. LIGO-III, EURO and EURO (xylophone), to probe
the interior of neutron stars. As clearly shown in Fig.~\ref{f2},
an inversion scheme using two leading QNM frequencies
 can readily discriminate between any two smooth EOSs considered
 there. The method
developed in the present paper in fact makes the construction of
such gravitational-wave detectors more rewarding.

\begin{figure}
\includegraphics[width=10cm,height=14.5cm]{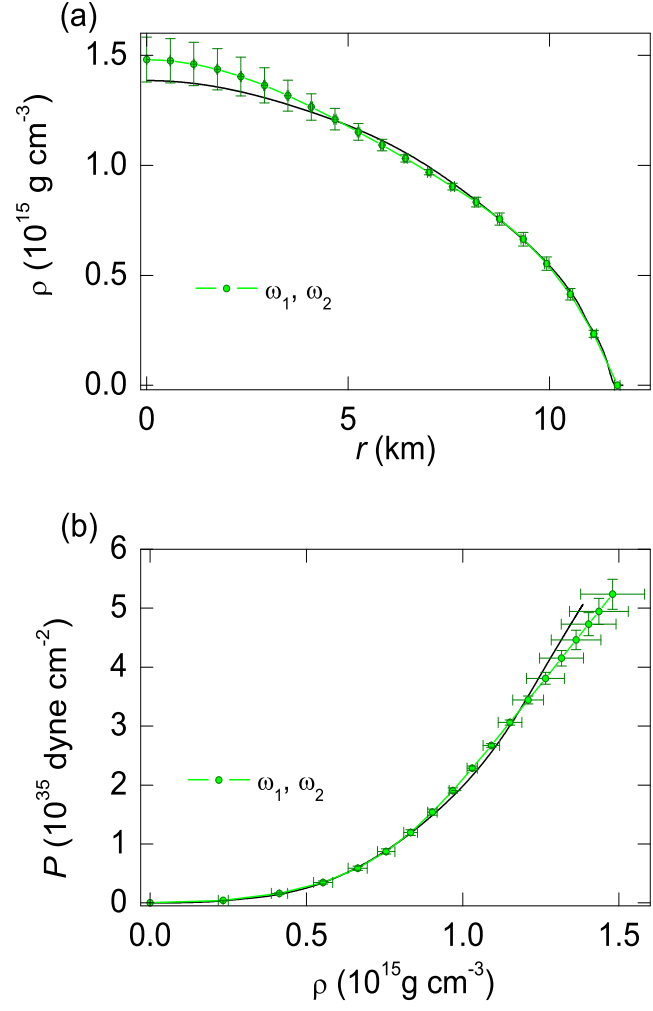}
\caption{Panels (a) and (b) depict $\rho(r)$ and $P(\rho)$
respectively for an APR1 star with $\cc=0.28$. The solid line is
the theoretical value, while the grey circles connected by a grey
line represent the results obtained from inversion scheme using
two approximately known leading axial $w$-modes and four
parameters. Twelve sets of data were used in the simulation. The
errors in the real and imaginary parts of the QNM frequency were
uniformly distributed in a $\pm 5\%$ interval. } \label{f12}
\end{figure}

In reality, gravitational signals are inevitably contaminated by
noises of various kinds, including thermal, quantum  and
gravity-gradient noises (see, e.g.
\cite{Liu:2000ac,Thorne:1998hq,Santamore:2001nr,Kokkotas_2001,GWQND_T,GWQND}
and references therein). Analogous to other fields of astronomy,
observational errors introduced by noises are deemed the major
hurdle to the implementation of the inversion scheme proposed
here. In order to infer the fine internal structure of neutron
stars, theoretical and experimental efforts must be done to
minimize the noise level, and sophisticated data acquisition
skills have to be developed to distinguish signals from noises
\cite{GWQND_T,GWQND,Cardiff}.  On the other hand, as demonstrated
in
 Fig.~\ref{f12}, our method remains accurate and stable against
 contamination of data created by noises.
The construction and operation of advanced gravitational-wave
detectors that can provide higher detection sensitivities and
resolution power will strongly boost the progress in
gravitational-wave asteroseismology \cite{GWQND_T,GWQND,Cardiff}.
The proposal in the current paper will then become a powerful
tool. As predicted in Refs.~\cite{super_prl,Cardiff}, the prospect
of gravitational-wave asteroseismology is bright.
\begin{acknowledgments}
We thank K Young,  WM Suen and LM Lin for discussions. Our work is
supported in part by the Hong Kong Research Grants Council (Grant
No: 401905) and a direct grant (Project ID: 2060260) from the
Chinese University of Hong Kong.
\end{acknowledgments}
\appendix
\newpage
\newcommand{\noopsort}[1]{} \newcommand{\printfirst}[2]{#1}
  \newcommand{\singleletter}[1]{#1} \newcommand{\switchargs}[2]{#2#1}

\newpage

\begin{table}
  \centering
  \begin{tabular}{|c|c|c|c|}
    \hline
    EOS &  $M$ ($10^{33}$ gm) & $R$ (km) \\
    \hline \hline
    APR1 & 4.460/4.475/4.460 & 11.83/12.12/11.70 \\
    \hline
    APR2 &  4.173/4.184/4.173 & 11.07/11.26/10.91 \\
    \hline
    AU & 3.843/3.862/3.843 & 10.19/10.61/10.08 \\
    \hline
    A & 3.289/3.282/3.289 & 8.723/8.529/8.564 \\
    \hline
    C & 2.791/2.788/2.791 & 7.404/7.185/7.216 \\
    \hline
    UT & 3.654/3.651/3.655 & 9.691/9.745/9.664 \\
             \hline
  \end{tabular}
  \caption{The mass $M$ and the radius $R$ of neutron stars constructed with
  different EOSs are compared with the corresponding values obtained from
  inversion using one or two QNMs. In each entry the values are
  obtained from TOV equation/inversion using $\omega_1$/inversion
  using $\omega_1$ and $\omega_2$ respectively.
  }
  \label{t1}
\end{table}
\begin{table}
  \centering
  \begin{tabular}{|c|c|c|c|c|c|c|}
    \hline
    Detector &  $f\,({\rm kHz})$ & $\tau \,({\rm ms})$ & $[S_n(f)]^{1/2}\,({\rm
    Hz}^{-1/2})$ & $(S/N)^2$ & $\sigma_f/f$ & $\sigma_\tau/\tau$ \\
    \hline \hline
    LIGO II & 7.320 & 0.065 & $1.13 \times 10^{-22}$ & $3.76 \times 10^{-1}$
    &$0.99$ & $3.8$\\

     & 12.76 & 0.037 & $1.98 \times 10^{-22}$ & $4.06 \times 10^{-2}$
    &$3.07$ & $11.6$\\
    \hline
    EURO  &  7.320 & 0.065 & $8.42 \times 10^{-25}$ & $6.80 \times 10^3$
    &$7.4 \times 10^{-3}$ & $2.8 \times 10^{-2}$ \\

     &  12.76 & 0.037  & $1.46 \times 10^{-24}$ & $7.45 \times 10^2$
    &$2.3 \times 10^{-2}$ & $8.5 \times 10^{-2}$ \\
    \hline
    EURO  & 7.320 & 0.065 & $4.93 \times 10^{-27}$ & $1.99 \times 10^8$
    &$4.3 \times 10^{-5}$ & $1.7 \times 10^{-4}$ \\
    (xylophone) & 12.76 & 0.037 & $2.83 \times 10^{-27}$ & $1.99 \times
    10^8$  &$4.4 \times 10^{-5}$ & $1.7 \times 10^{-4}$ \\
             \hline
  \end{tabular}
  \caption{The noise amplitude spectral density $[S_n(f)]^{1/2}\,({\rm
    Hz}^{-1/2})$, the signal to noise ratio $(S/N)^2$,  $\sigma_f/f$ and
     $\sigma_\tau/\tau$ are tabulated for LIGO II, EURO  and
     EURO  (xylophone) gravitational-wave detectors.
     Here the frequency and damping time of the two leading axial $w$-modes of an APR1 star with
     $\cc=0.28$ are considered, which are also used as the input in
     Fig.~\ref{f12}, and $E_{\rm gw} = {10^{-6}{\rm M}_\odot c^2}$ and
${r} = 10\, {\rm kpc}$ are assumed.
  }
  \label{t2}
\end{table}
\clearpage
\newpage

\newpage 

\end{document}